# Relaxation of interstitials in spherical colloidal crystals

D. S. Roshal, A. E. Myasnikova, S. B. Rochal

Faculty of Physics, Southern Federal University, 5 Zorge str., 344090, Rostov-on-Don, Russia

Graphical abstract:

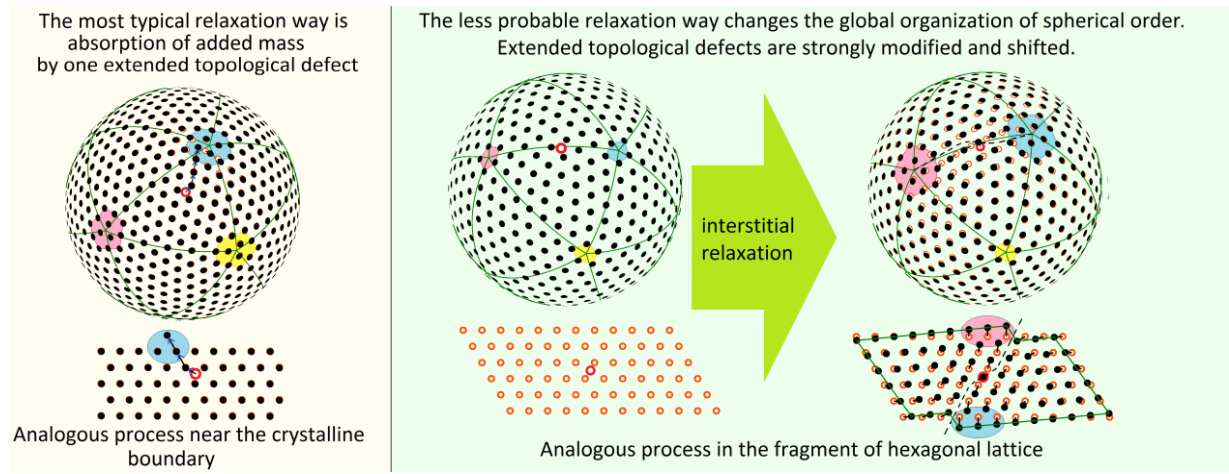

*Highlights:*

- Ways of interstitials relaxation in spherical colloidal crystals are classified.
- The most typical relaxation way involves only one extended topological defect.
- Parent phase reveals changes in global organization of relaxed hexagonal order.
- Interstitials fractionalization is analyzed in terms of parent icosahedron method.
- Relationships between relaxation in planar and spherical crystals are found.

Spherical colloidal crystals (CCs) self-assemble on the interface between two liquids. These 2D structures unconventionally combine local hexagonal order and spherical geometry. Nowadays CCs are actively studied by altering their structures. However, the statistical analysis of such experiments results is limited by uniqueness of self-assembled structures and their short lifetime. Here we perform numerical experiments to investigate pathways of CC structure relaxation after the intrusion of interstitial. The process is simulated in the frames of overdamped molecular dynamics method. The relaxation occurs due to interaction with extended topological defects (ETDs) mandatory induced in spherical CCs by their intrinsic Gaussian curvature. Types of relaxation pathways are classified and their probabilities are estimated in the low-temperature region. To analyze the structural changes during the relaxation we use a parent phase approach allowing us to describe the global organization of spherical order. This organization is preserved by only the most typical relaxation pathway resulting in filling one of vacancies integrated inside the ETD areas. In contrast with this pathway the other ones shift the ETDs centers and can strongly reconstruct the internal structure of ETDs. Temperature dependence of the relaxation processes and the mechanism of dislocation unbinding are discussed. Common peculiarities in relaxation of spherical structures and particular fragments of planar hexagonal lattice are found.

## 1. Introduction

The fact that colloidal particles can be adsorbed at the interface between two liquids has been known since the beginning of the 20[th] century [1]. However, the first colloidal crystals on non-planar surfaces were obtained only recently [2]. These 2D objects are characterized by a local hexagonal order and the obligatory presence of topological defects induced by non-zero Gaussian curvature [3-5]. Topological defects occupy the extended areas on the surface of colloidal crystals (CCs). In these regions a local hexagonal order is significantly distorted. If the defect area has non-zero topological charge [4] the defect is called a scar [4, 6], otherwise a pleat [7-8]. The presence of 12 scar-like extended topological defects (ETDs) is typical of the spherical CCs [3-6]. The centers of these ETDs are approximately equidistant from each other, so they are located near the vertices of the icosahedron [3]. Pleats are usually observed in the CCs which are self-assembled on the surfaces with negative Gaussian curvature [7-9].

Over the last decade experimental methods for studying 2D CCs were substantially advanced. Previously, the CC structures were simply observed with an optical microscope [2, 4]. Now it is possible to study these objects by active altering their structure with the new experimental methods, which originate from the optical tweezers technique. Use of these methods allows moving individual colloidal particles [8, 10] or coherent changing the positions of the whole groups of them [8, 10]. After the enforced reconstructions colloidal structure relaxes. This process can be directly observed and registered using an optical microscope.

It is especially interesting to observe the relaxation of the interstitials intruded into spherical CCs [8], where the explicit crystal boundary is absent. After the intrusion of an extra particle, the local hexagonal order is restored around it and one or few nearest ETDs are reconstructed. As is well known, the interstitial close to the crystal boundary can relax due to a shift of lattice nodes towards the crystal boundary [11-13]. In this case the last particle of the shifted line is pushed upon the crystal surface. Such a process can be considered as a movement of two dislocations [11-13] with zero total Burgers vector to the crystal boundary. Relaxation of the interstitial in the spherical CCs occurs similarly since the latter process also represents the movement of a few dislocations [8, 14], which are absorbed by the closest ETDs. The well-known interpretation of scars as grain boundaries [4] allows us to understand that the dislocation absorption by scar is analogous to the process of its absorption by the crystal surface [8]. In both cases, the dislocation charge is not conserved [15].

The crystal boundary can absorb an unlimited number of dislocations. In contrast, the scar is always finite, and can be always characterized by a small *discrete* number of vacancies embedded into its structure [15]. Nevertheless, during the relaxation the interstitial can be '*fractionalized*' [8]. In other words the additional mass of the single intruded particle is redistributed over the spherical crystal. Here we study in details this process and show that such mass splitting is only one side of the interstitial relaxation process. The other side is the change in the global organization of spherical hexagonal order. This organization is analyzed in terms of a parent icosahedron method that is proposed in this article.

There are a large number of works devoted to the relaxation of interstitials and vacancies in flat [8, 16] and spherical [8, 10, 14] crystals. However, as far as we know, the defect relaxation has never been considered in terms of preserving or changing the global organization of the hexagonal order on the sphere. Here we develop the parent phase approach and demonstrate how the global order reconstruction occurs while several ETDs simultaneously participate in the relaxation of interstitial. The article also clarifies analogies between the relaxation processes occurring in spherical CCs and those, which take place in the finite fragments of an ordinary 2D hexagonal lattice.

## 2. Theoretical methods

In this section we recall some known properties of the low-energy spherical hexagonal order. To characterize the processes of the interstitial relaxation we introduce a concept of global organization of such spherical order. After that we consider the overdamped dynamics of colloidal particles and apply it to model the interstitial relaxation.

### 2.1 Low-energy hexagonal order on the sphere and the parent icosahedron approach

The particles in the colloidal crystal are retained by surface tension at the interface between the two liquids and interact each with other with screened Coulomb potential [8]. Ordinarily the interaction potential is simplified and replaced by conventional Coulomb-like expression corresponding to the particles repulsion. Thus, the order in the spherical CCs is successfully modeled and analytically studied [17-19] in the frame of the simplest electrostatic energy:

$$U = \sum_{j>i}^{N} \frac{1}{r_{ij}}, \qquad (1)$$

where $r_{ij}$ is the distance between *i-th* and *j-th* particles, *N* is the total number of particles. The self-assembly of 2D structure on a curved surface is described by the conditional minimization of the energy (1) with respect to coordinates of the particles. The condition imposed is that any particle during minimization should be retained on the surface under consideration.

Structures that for a given N correspond to the global minimum (1) are the solutions of the Thomson problems [20-24]. The equilibrium energies (1) of structures corresponding to global and local-minima are very close. Moreover, the difference between the equilibrium energies (1) is strongly reduced and the number of equilibrium structures grows exponentially with the number *N* of particles in the structure [25]. Therefore, the structure of the real CC would rather correspond to one of the numerous deep local minima of energy (1), than to exact solution of the Thomson problem. However, since the solutions of Thomson problems are well known from the literature it is more convenient to use them to model the order in spherical CCs. Therefore here we study the interstitial relaxations on examples of model spherical structures (MSSs) corresponding to the deepest known today minima of energy (1).

Now let us recall some unusual properties of the hexagonal order on the sphere [5, 13]. Sufficiently large lowest energy MSSs with N≥400 have 12 scar-like ETDs [3, 23] which are located near the vertices of regular or slightly distorted spherical icosahedron covered by a simply connected hexagonal lattice. Formation of ETDs in the spherical crystals is induced by the Gaussian curvature of the spherical surface. The hexagonal order in the defect areas is strongly distorted but any defect can be surrounded by a pentagonal contour with sides being parallel to the minimal translations of the order [15]. The pentagonal shape of the surrounding contour [5, 15] is caused by the fact that each ETD has the minimal positive total topological charge +1.

Below we apply a parent icosahedron (PI) method to characterize the global organization of the hexagonal order on the sphere. So let us describe the PI construction briefly. Triangulation of the order inside the ETD reveals the linear scar [4], which is a chain of alternating particles with 5 or 7 nearest neighbors. Virtual insert of a small number $N_{ins}$ of particles inside the pentagonal contour surrounding the ETD allows the topological reconstruction of the order inside the defect [15]. In the result of this procedure all particles inside the defect area except the single one obtain six neighbors and the scar disappears [15]. Location of the appearing 5-fold disclination is completely determined by the location and shape of the pentagonal contour surrounding the initial ETD [15]. Twelve local 5-fold disclinations appear on the sphere after applying the above procedure to all the ETDs. These disclinations are located at the vertices of spherical icosahedron, which is smoothly covered by the hexagonal order. We call the resulting spherical structure the parent icosahedron. Let us stress that the PI can be irregular. Similar PIs have been recently used to advance the Thomson problem [24]. As far as we know, for the first time the analogous regular spherical icosahedra were used by Caspar and Klug to construct their model of spherical viral capsids [26].

Idea of PIs develops the parent phase approach, which is fruitfully used in many fields of physics during the last 6-7 decades. See, for example [27-28]. PIs are also very important for our approach, because any low-energy MSS can be presented as a result of exclusion of some nodes located near the vertices of the corresponding PI. Therefore, on the one hand the PI geometry characterizes the global organization of the hexagonal order on the sphere, and on the other hand the PI allows us to characterize each ETD in the MSS by the particular number $N_{ins}$ of vacant positions. Note also, that the 12 ETDs are the only possible defects in the low-energy MSSs, where all dislocations are absent. The dislocations are possible only in spherical structures with the energies that are not close to the deepest minimum of energy (1).

## 2.2 Interstitial relaxation in the frames of overdamped dynamics

To study the relaxation of interstitial it is enough to intrude an additional particle at a certain position on the sphere, and then to apply a simplified molecular dynamics approach described below. In viscous medium the particle motion equation reads:

$$m\mathbf{a}_i \Big|_{\|} = \left( -\eta \mathbf{v}_i + \mathbf{F}_i(T,t) - \frac{\partial U}{\partial \mathbf{r}_i} \right)\Big|_{\|}, \qquad (2)$$

where $i=1,2...N$; m, $\mathbf{a}_i$, $\mathbf{v}_i$ and $\mathbf{r}_i$ are the mass, acceleration, velocity and radius-vector of the $i$-th particle, respectively. Coefficient $\eta$ stands for the viscous friction of the particle. Random in time $t$ force $\mathbf{F}_i(T,t)$ describes the Brownian motion of colloidal particles. Its amplitude increases with the temperature $T$ growth. Since the particles are retained on the spherical surface only the parallel (tangent) projection of Eq. (2) is taken into account to describe the particle motion.

The mass $m$ and friction coefficient $\eta$ for colloidal particle are proportional to the third and first degrees of the particle size which is about several microns. Therefore with the decrease of the particle size its mass $m$ decreases much more quickly than the viscous friction. We assume that like the case of other micron-size systems moving in viscous liquid (lipid membranes, vesicles of different kinds) [29] the dynamics of colloidal particles is overdamped and solve numerically the system of following simplified motion equations:

$$\eta \frac{\partial \mathbf{r}_i}{\partial t}\Big|_{\|} = \left( \mathbf{F}_i(T,t) - \frac{\partial U}{\partial \mathbf{r}_i} \right)\Big|_{\|}. \qquad (3)$$

After the small time interval $dt$ the particle shift $dr_i$ is determined as:

$$dr_i\Big|_\| = \frac{dt}{\eta}\left(\mathbf{F}_i(T,t) - \frac{\partial U}{\partial \mathbf{r}_i}\right)\Bigg|_\|. \tag{4}$$

At low temperature ($\mathbf{F}_i(T,t) \approx 0$) numerical solution of Eqs. (3) is equivalent to minimization of energy (1) with the conventional gradient descent method. Of course, this minimization is conventional because during the process the particles are retained on the spherical surface.

Another interesting limiting case (implemented in our model) takes place for the structure considered near the equilibrium, where

$$\frac{\partial U}{\partial \mathbf{r}_i}\Bigg|_\| \approx 0. \tag{5}$$

Then the main contribution to the left part of Eq. (4) arises from random Brownian forces $\mathbf{F}_i$ which act on the particles during short time intervals $\Delta t_i$ located randomly on the time axis. Near the equilibrium (due to Eq. (5)) the effect of Brownian forces simply leads to a shift $\Delta \mathbf{r}_i\big|_\|$ of the particle:

$$\Delta \mathbf{r}_i\Big|_\| \approx \frac{1}{\eta}\int_{\Delta t_i} \mathbf{F}_i\Big|_\| dt. \tag{6}$$

We assume that the shifts (6) occur instantly ($\Delta t_i$ tend to zero) and use the simplified equation:

$$d\mathbf{r}_i\Big|_\| = \frac{dt}{\eta}\left(\frac{\partial U}{\partial \mathbf{r}_i}\right)\Bigg|_\| + \varsigma_i(t)\Big|_\|, \tag{7}$$

where functions $\varsigma_i(t)\big|_\|$ are equal to zero everywhere except the random time moments in which they take the random values (6).

Let us consider the numerical solution of Eqs. (7) in more details. First of all we put $\eta = 1$ (the reason of this simplification is explained below) and set the value of an average time $t_{av}$ between Brownian impacts on the particles. The algorithm works step by step. In course of *j*-th step for each particle we calculate its shift (7). To calculate the random part of this shift we use a conventional white random function *rnd* (0<*rnd*<1) which present in any programming language. Let the *j*-th step corresponds to a time interval $dt_j$. The ratio $dt_j/t_{av}$ is equal to probability that the Brownian impact on the particle occurs during the *j*-th time interval. Thus if the ratio $dt_j/t_{av}$ is smaller than the current *rnd* value generated for *i*-th particle we put the random shift of the considered particle to be equal to zero: $\varsigma_i(t_j)\big|_\| = 0$. Else we calculate this shift value in the frame of assumption that the shift is described as a white random function with some maximal amplitude *A*.

In the presented algorithm the change of viscosity $\eta$ is equivalent to the time scaling with the appropriate variation of the ratio $t_{av}/dt$ (see Eq. (7)). Therefore, till we do not consider the relaxation time of spherical structure, our choice $\eta = 1$ is justified. Thus the relaxation process in our model is governed by only two parameters. They are the amplitude of random shifts *A* and the ratio $t_{av}/dt$. Both of these parameters affect on the system temperature *T*. Growth of the amplitude *A* increases the temperature while growth of the $t_{av}$ value decreases it.

Due to the thermal fluctuations of particles positions the energy (1) increases with respect to its minimal value in the static equilibrium. The temperature $T$ of the structure can be expressed in terms of the resulting increment $\Delta U$ of the energy (1). In the model, where the particles are retained on the spherical surface, one can consider only in-plane (parallel to the sphere surface) fluctuations of the particles positions. The mean-square fluctuations $<\Delta r_i^2>$ are proportional to the temperature and satisfy the equipartition theorem even in the case of overdamped dynamics. The structure with $N$ particles has $2N$ degrees of freedom. Each degree of freedom corresponds to the average fluctuation energy $\frac{1}{2}k_B T$. Thus, near the equilibrium the average temperature can be calculated as:

$$<T> = \frac{<\Delta U>}{N k_B}, \qquad (8)$$

where $<\Delta U>$ value is deduced numerically by $\Delta U$ averaging over a period of time much greater than $t_{av}$ value. Since the energy (1) in the model is defined up to unessential coefficient (which is chosen to be equal 1) the temperature (8) in the model is also defined up to the same coefficient. Therefore it is reasonable to estimate the temperature in terms of a ratio between the mean value of $\sqrt{<\Delta r_i^2>}$ and the average distance between nearest particles. The fluctuations are important in the model since they can prevent the relaxing system from a fall into not deep minima of energy (1). In other words, sufficiently high temperature prevents freezing of dislocations during their motion toward the topological defects.

## 3. Model spherical structures

We intruded interstitials and considered their relaxations and reactions with ETDs in about twenty MSSs, which were either generated by us [24], or found in the literature. The right column of fig. 1 shows two examples of the considered MSSs. The first of these structures [24] (N = 657) (see fig. 1 (a)) represents the most common type of Thomson problem solutions. In similar Thomson structures ETDs appear due to exclusion of nodes located in the vertices of the PI [24]. Each ETD of such type can absorb only one particle without global reconstruction of the spherical order. The second structure, shown in fig. 1 (c), is also potential solution (N = 912) [22] of the Thomson problem. However, unlike the first MSS, the second one is characterized by more complex ETDs of different types. Here ETDs occupy relatively large areas of the structure surface and contain from one to three vacant positions. Due to the greater complexity of ETDs the second MSS is closer to the experimentally observed structures of spherical CCs.

Panels (b) and (d) (see fig. 1) demonstrate the PIs for Thomson structures shown in panels (a) and (c), respectively. In the panels' centers the ETDs areas are highlighted by dark shades of blue, yellow and red colors. The pentagonal contours surrounding the defects are equilateral, with side lengths equal to 1 and to 2 for the MSSs, shown in fig. 1 (a) and fig. 1 (c), respectively. A choice of ETDs boundaries in the form of equilateral pentagons simplifies the construction of PIs, because in this case the centers of pentagonal contours coincide with the PI vertices. To idealize the order in the first MSS and thus to obtain the first PI one particle is inserted within each pentagonal boundary. In fig. 1 (b) the inserted particles are shown by open circles. In the visible part of the second MSS (fig.1 (c, d)) there are three ETDs with three vacancies in each defect. Inserted particles as well as the particles present in the defect areas before the order restoration are shown in fig. 1 (d) with open circles. Obtained PIs are unambiguously characterized by triangulation indices of their edges, which in turns determine the global organization of the spherical hexagonal order. Let us recall that the edge triangulation indices are calculated from a triangle with the longest side coinciding with this edge. Triangulation indices of the edge are equal to the lengths of two other sides of the triangle provided these sides are parallel to the minimal translations of hexagonal order. In fig. 1 (b, d) the mentioned triangles are highlighted in grey, triangulation indices are (4, 6) and (8, 4), respectively.

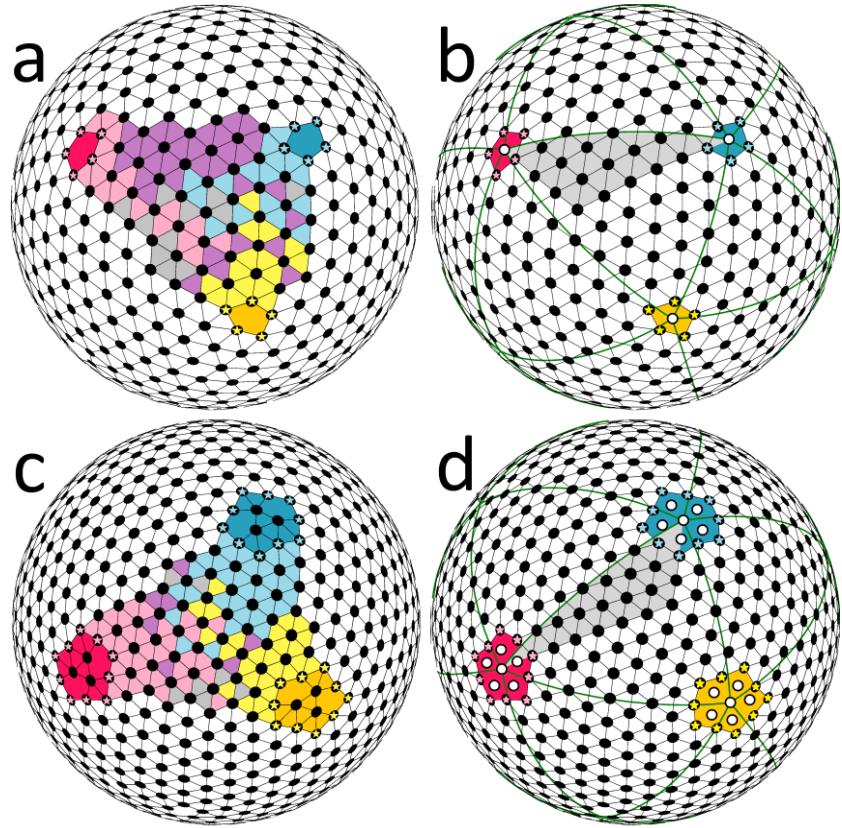

*Figure 1. Global organization of the spherical hexagonal order and different pathways of the interstitial relaxation. Panels (a, c) show the low-energy model spherical structures. Panels (b, d) demonstrate PIs for the model structures (a, c), respectively. The PI is obtained by inserting some additional nodes into the areas of defects. Regular or irregular PIs are always covered with the maximally perfect hexagonal order possible on the sphere: all nodes except the twelve ones at the PI vertices have 6 neighbors. We consider the structures (a, c) as models of the order in spherical colloidal crystals and use them to study the interstitial relaxation. Interstitials were intruded subsequently between three neighboring particles lying at the vertices of all colored triangles. Relaxed structures were obtained by the energy (1) minimization. Different colors of triangles on panels (a, c) correspond to qualitatively different pathways of the relaxation process occurring in the low-temperature limit. If the particle is intruded into any blue, pink or yellow triangle then relaxation results in filling a vacancy inside the dark blue, dark pink or dark yellow ETD, respectively. Relaxation of interstitial intruded into any purple triangle shifts ETDs and changes their shape. In the case when the particle is intruded into grey triangle the frozen dislocation remains in the structure after relaxation.*

### 4. Results of numerical experiments and their discussion

During our numerical experiments we intruded an additional particle into the MSS in all positions located between three nearest neighbors in the centers of all different triangles obtained from the conventional triangulation [4] of the MSS. Each time after the intrusion we studied the structure relaxation. We found that at zero temperature in all the considered MSSs all relaxation processes are univocally determined by only the initial spherical structure organization and the initial coordinates of the intruded particle. Even a temperature growth leading to thermal fluctuations up to 15-20 percent of the average distance between the particles does not affect qualitatively on most of the relaxation processes. Further temperature growth induces a dynamic disorder of particles in spherical structures and destroys the structural relations between the relaxed and initial structures.

We classify the numerous similar processes in a few different pathways which are discussed below. The percentage of any relaxation pathway $P_j$ can be calculated as

$$P_j = \Delta M_j / M_0 , \qquad (9)$$

where $\Delta M_j$ is the number of the triangles for which the particle intrusion induces the structure relaxation according to $j$-th pathway, and $M_0$ is the total number of triangles resulted from the structure triangulation. Since all triangles have approximately equal areas Eq. (9) can be rewritten in another equivalent form:

$$P_j \approx \Delta S_j / S_0 , \qquad (10)$$

where $\Delta S_j$ is the total area of the triangles for which the particle intrusion induces the structure relaxation according to $j$-th pathway, and $S_0$ is the total area of the spherical structure. Since some of the relaxation processes depend on the temperature and lead to different relaxed structures after some temperature growth, the value $\Delta M_j$ as well as the percentage (9-10) are also changed. However, as soon as the thermal fluctuations overcome 5-10 percent of the average distance between the particles the switching between pathways is stopped. Note also that if during an experiment the particle is intruded with equal probability in different points over the MSS surface then Eqs. (9-10) yields the probability to observe the given relaxation pathway.

Let us start from the relaxation pathway with the maximal $P_j$ value. Our numerical simulation shows that the large number of vacant positions in ETD located near the interstitial increases the percentage of relaxation pathway which preserves the PI structure and is finished with filling a vacancy in a single ETD. For the structures shown in fig.1 (a, c) this-type relaxation takes place if a particle is intruded in the center of yellow, blue or pink triangles. For the sake of clarity only the triangles in the region between three ETDs located in the center of panels (a, c) are colored. This coloring corresponds to the zero temperature limiting case. During the relaxation process the interstitial induces the consecutive particles' shift along a polyline with links being the minimal translations of the hexagonal order. The last particle of this polyline is pushed into the ETD colored a bit darker than the triangle where the particle was intruded (see fig.1 (a, c)). Figure 2 (a) shows how this relaxation proceeds in the structure shown in fig. 1 (c). Since the ETDs of the structure shown in fig.1 (c) contain more vacant positions, the percentage of this relaxation pathway for the structure (c) is higher than for the structure (a).

Thus, for the structure (c) the percentage (9-10) of the discussed above relaxation process, which leads to the absorption of added mass by one ETD, is greater than 0.8, while for the structure (a) the analogous percentage is less than 0.5. For all considered MSSs with $N<1000$ this relaxation process is temperature independent. To conclude this pathway discussion let us stress, that the relaxation of an interstitial inserted near the boundary of the plane hexagonal lattice always occurs in a similar way (see Fig. 2 (d)). This figure as well the figs. 2 (e, f) were obtained by numerical simulation of the relaxation process in the shown fragments of the plane hexagonal lattice. To construct them the Lennard - Jones potential [30] was used to model the interaction between particles and the temperature was taken to be zero.

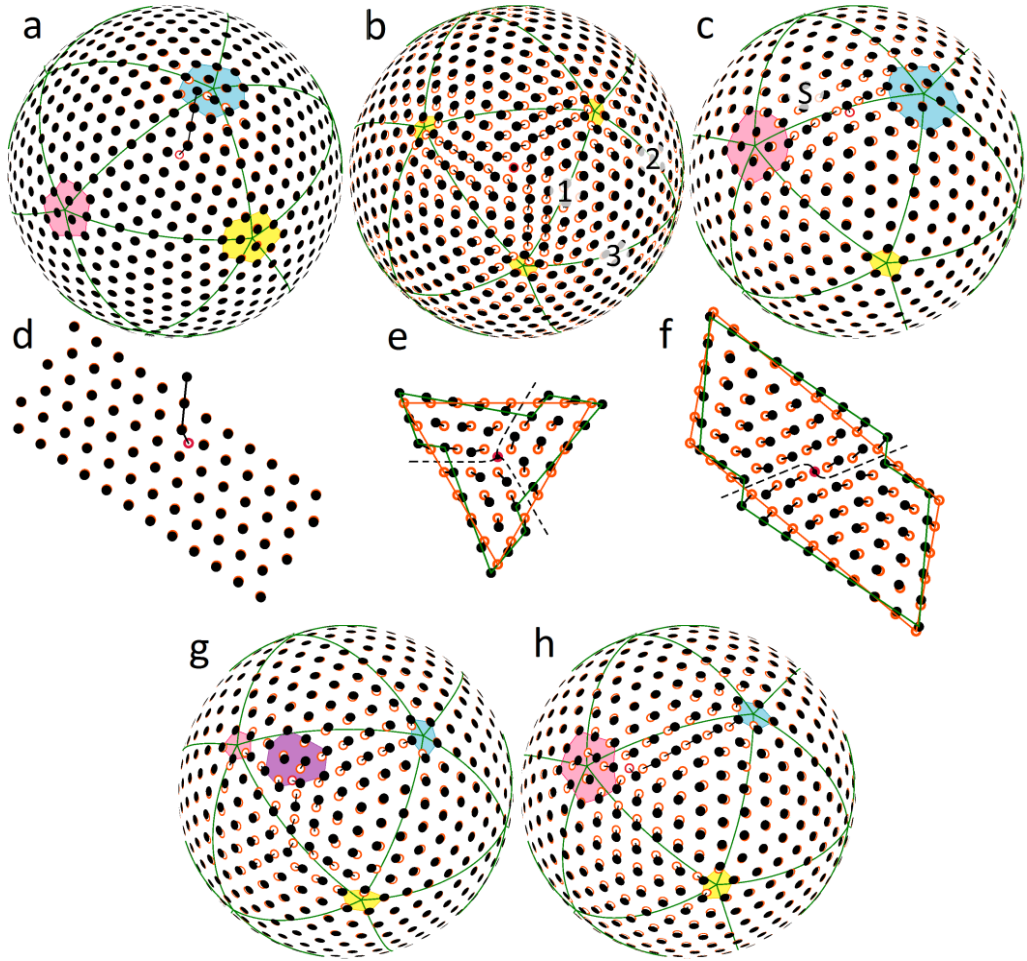

*Figure 2. Mechanisms of the hexagonal order relaxation after the intrusion of interstitial. The initial positions of particles are shown by small orange open circles. The intruded particle is presented by big red open circle. The particles after relaxation are shown by black circles drawn over the orange ones. Panel (a) shows the most probable way of interstitial relaxation in the colloidal crystal. The interstitial shifts the chain of particle towards the nearest ETD and the closest to ETD particle is absorbed by it. Only this relaxation mechanism preserves the PI and, consequently, the global organization of the spherical order. Panel (b) shows an example of relaxation pathway altering the PI. Relaxation of Caspar and Klug icosahedron (3, 8) takes place after the intrusion of a particle in the center of regular triangular face. This relaxation was considered in Ref. [14, 31], but reconstruction of the global organization of spherical hexagonal order during the process was not discussed. During the pathway (c) the interstitial shifts the particles in opposite directions towards two nearest ETDs. One can note that the PI edge connecting the centers of two involved ETDs is increased. (d-f): Interstitial relaxations similar to pathways (a-c) take place in appropriate fragments of plane hexagonal lattice. In cases (e, f) the relaxation results in the relative shift and turn of lattice parts separated by dotted lines. (g): Rarely, one of the dislocations arising due to the interstitial intrusion is stopped outside the ETD area. The frozen dislocation is surrounded by a purple hexagonal scalene contour. The resulting structure is very resistant to thermal fluctuations. (h): Only the growth of their amplitude up to 10 per cent of the average distance between particles brings the dislocation into the motion, which is finished with the dislocation absorption by the second (pink) ETD.*

    Now we proceed to other relaxation pathways, which mandatory alter the global organization of the spherical order. Let us stress that the MSS, where the topological defects are

the local disclinations, in principle cannot relax with preservation of its global organization (see fig. 2(b)). The absorption of dislocation by the local disclination necessarily results in formation of ETD with the center shifted from the position of initial disclination. This fact can be demonstrated on an example taken from Ref. [14, 31], which considers relaxation of the interstitial intruded exactly in the center of a regular triangular face of the Caspar and Klug icosahedron (3, 8). Interstitial pushes particles around itself simultaneously in three directions, which corresponds to the motion of three dislocations. Each of them moves to the corresponding disclination and, finally, the dislocations are absorbed (compare fig. 2 (b) and 2 (e)). Let us consider how this process changes the PI. After the relaxation each appeared (yellow) ETD has a single vacancy (see fig.2 (b)) and the distance between the appeared ETDs is characterized with indices (4, 7). It is easy to see that the distance between the defects is slightly increased with respect to the edge length of the initial regular icosahedron: $4^2+4\cdot7+7^2 > 3^2+3\cdot8+8^2$. This reconstruction of the global spherical hexagonal order preserves the three-fold axis. However, the lengths of the edges indicated in Fig. 2(b) with the numbers 2 and 3 also vary and after the relaxation these lengths are determined by triangulation indices (8, 2) and (9, 3), respectively. This relaxation pathway is also independent on temperature.

Note, that although the symmetric pathway proposed in Ref. [14, 31] is very nice from a mathematical point of view, in the frames of our model we have never identified the process like this in the MSSs, which were more or less similar to real CCs. We assume that the symmetric pathway proposed in Ref. [14, 31] is possible only in the high-symmetry MSS and prohibited in the real asymmetric colloidal crystal, where the local disclinations located at the vertices of an equilateral triangle are absolutely improbable. However, only the further experimental research can answer the question whether such relaxations are possible in spherical colloidal crystals.

Let us proceed further and discuss two more probable relaxation pathways. Intrusion of a particle into purple triangles (see fig. 1(a,c)) results in an independent on temperature process, which is finished by absorption of emitted dislocations by two different ETDs (see fig. 2 (c)). Similar process of interstitial relaxation occurs in the narrow strip of hexagonal lattice shown in fig. 2 (f), where two emitted dislocations are absorbed by different sides of the strip. Fig. 2 (c) demonstrates that the distance S between the centers of two involved ETDs slightly decreases. In the initial structure (see fig. 1 (a)) the distance S is defined as: $S^2=4^2+6^2+4*6=76$. After relaxation of the interstitial this PI edge becomes shorter: $S^2=2^2+7^2+2*7=67$ (see fig. 2 (c)). The analysis of our numerical experiments, as well as direct comparison between fig. 1 (a) and fig. 1 (c) show that the percentage of such pathway decreases with an increase of the number of vacancies in ETDs which are contained in the MSSs.

Similarly, growth of number of vacancies located in ETDs decreases the percentage of the last relaxation pathway which can occur if the particle is intruded into triangles highlighted with gray in fig. 1 (a) and fig. 1 (c). In this case, the relaxation process is temperature dependent. At low temperatures, one of the dislocations emitted after the intrusion of interstitial freezes and does not reach the ETD area (see Fig. 2(g)). The structure obtained is not energetically favorable and the dislocation remains frozen up to temperatures corresponding to thermal fluctuations about 5 or 10 percent of the average distance between the particles. A further temperature increase unbinds the dislocation, and it is absorbed by the second ETD (which is pink in fig. 2(h)). Thus, the role of temperature in the process is only to overcome the potential barrier which stabilizes the metastable spherical structure with frozen dislocation. At a sufficiently high temperature processes shown in fig.2 (g) and (h) are qualitatively indistinguishable. Therefore we disagree with the statement [31] that unbinding of dislocations (which are induced due to the interstitial) is associated with the curvature of the structure. In our opinion the role of curvature is only to induce ETDs and the dislocation unbinding occurs due to two main reasons. The first of them is the temperature growth and the second one is the finite size of spherical crystals. Due to the finite size of MSS the dislocation always appears relatively close to one of ETDs and short-range interaction with this ETD leads to the dislocation absorption by the defect. In planar 2D crystals the temperature also easily unbinds the dislocations located near the crystal boundary.

The process shown in fig. 2(h) has one more interesting peculiarity. For the ETD, which is highlighted by blue in fig. 1(a) and 2(h), the absorption of dislocation corresponds to its simple rotation (compare orientations of the defect in fig. 1(a) and 2(h)). The single vacancy embedded in this defect is not filled. Therefore, in the considered case the added mass is also absorbed due to the global rearrangement of spherical order and it is not possible to say that the added mass is split and absorbed only by two ETDs, which participate in the process.

## 5. Conclusion

This work deals with the interstitials relaxation and underlying reactions between ETDs and interstitials intruded in the spherical colloidal crystals which are simulated by MSSs being the potential solutions of the Thomson problem [20-24]. In these model structures the low-energy hexagonal order smoothly covers the spherical surface except the areas of 12 curvature-induced ETDs. To characterize this order and interpret the results of our numerical experiments on interstitial relaxation we use the parent phase approach. We introduce the parent icosahedron (PI) structure, where the order inside ETDs is restored until a single local disclination is remained. The centers of ETDs in the initial MSS coincide with the PI vertices. The PI edges are the hexagonal order translations which in general case can have slightly different lengths. On the one hand the motion of ETDs centers during the interstitial relaxation changes these translations and indicates a reconstruction in the global organization of the spherical order. On the other hand the use of the PI allows characterizing each topological defect with some number of embedded vacancies that can be filled during the relaxation.

Suggested classification of the relaxation pathways and expression to calculate their percentage (Eq. (9-10)) enable us to establish that the most typical pathway of the interstitial relaxation preserves the global order organization. During such relaxation the interstitial induces the particles' shift along the polyline with links being parallel to the minimal translations of the hexagonal order. The last particle of this polyline is pushed into the ETD which does not move since it absorbs simultaneously two dislocations with zero total topological charge. Relaxation of additional particle intruded near the crystal boundary occurs in a similar way.

All other pathways are shown to be less typical. They induce the shifts of the ETDs centers and change the global order organization. During the most typical pathway from this series two ETDs are shifted and the added mass is distributed between them. Thus, our approach shades a new light on the problem of interstitial fractionalization [8]. We show that in the appropriate narrow fragment of the planar hexagonal lattice the interstitial is 'fractionalized' analogously. At low temperatures, this-type relaxation can be unfinished if one of the dislocations (emitted due to intrusion of interstitial) is bound before reaching the ETD. However, the temperature growth allows ending the process. Since in planar 2D crystals the temperature also easily unbinds the dislocations located near the crystal boundary we state that the main mechanism which unbinds the dislocations in the spherical structure is associated with the finite size of the system. The curvature only induces ETDs and due to the finite size of spherical crystal the dislocation always appears relatively close to one of ETDs which play the role similar to the boundary of planar crystal.

In future experimental works it may be interesting to check our theoretical findings. Also it is worth to investigate with both theoretical and experimental methods the relaxation of vacancies induced in spherical colloidal crystals. As far as we know this process is still uninvestigated in spite of the well-known fact that in solid crystals the diffusion of vacancies and the motion of interstitials occur in different ways. Our results may be interesting for physicists working on theoretical and experimental problems of self-assembly and enforced structure reconstruction in various types of spherical nano-and micro-structures.

## Acknowledgements

Authors acknowledge financial support of the RFBR grant 13-02-12085 ofi_m.